\documentclass[a4paper,11pt]{article}
\pdfoutput=1 

\usepackage{jcappub}
\usepackage{aas_macros}
\usepackage{amsmath}
\usepackage{natbib}

\usepackage[T1]{fontenc}

\title{\boldmath Robustness of Solutions of the Quantum Kinetic Equations in the Presence of Matter Density Fluctuations}

\author[a]{Shashank Shalgar}
\author[a]{and Mariam Gogilashvili}

\affiliation[a]{Niels Bohr International Academy and DARK, Niels Bohr Institute, University
of Copenhagen, Blegdamsvej 17, 2100, Copenhagen, Denmark}

\emailAdd{shashank.shalgar@nbi.ku.dk}
\emailAdd{mariam.gogilashvili@nbi.ku.dk}

\abstract{We investigate the role of fluctuations in the matter density on neutrino flavor evolution by studying their effects on the collision terms in the spherically symmetric quantum kinetics equations (QKEs). We solve the QKEs with varying radial resolution ($r_{\mathrm{bins}} = 150 \, , 1500 \, , 15000$) to assess numerical convergence in angular distributions, number densities, and energy spectra for four neutrino flavors ($\nu_e$, $\bar{\nu}_e$, $\nu_x$, $\bar{\nu}_x$). Our results demonstrate that the solutions are numerically converged already at the coarsest resolution, with higher resolutions yielding almost identical outcomes. We introduce random perturbations to each radial bin, thus adding perturbations with a length scale that is related to the radial resolution. We study both time-independent and time-dependent perturbations to the matter density that affect the collision term and analyze their effects on neutrino flavor evolution. We find that such fluctuations do not induce any significant instabilities or qualitative changes in flavor evolution. Angular structure remains robust, and flavor-dependent number densities and energy spectra show only minor deviations compared to the unperturbed case. These findings suggest that matter perturbations have a negligible effect on neutrino flavor evolution in spherically symmetric settings.}

\begin{document}
\maketitle
\section{Introduction}
Core-collapse supernovae (CCSNe) are among the most energetic events in the universe, powered largely by the intense neutrino flux emitted during stellar collapse and the subsequent cooling of the proto-neutron star. Neutrinos carry away the vast majority of the gravitational binding energy and play a central role in the dynamics of the explosion and in setting the conditions for nucleosynthesis \cite{Colgate:1966ax, Bethe:1985sox}. Despite decades of research \cite{Bethe1990, BURROWS1995, Janka1996, JANKA2001, Burrows2012, Muller2016, Radice2016}, the exact role of neutrinos in powering the explosion remains an open question, particularly when it comes to the effects of flavor evolution \cite{Mirizzi:2015eza, Duan:2010bg}.

In regions of high neutrino density, the coherent forward scattering of neutrinos due to other neutrinos in the medium cannot be ignored~\cite{Duan2006, Duan2006b, Duan:2010bg}. Although the neutrino number density is the largest in the core of the proto-neutron star, the existence of flavor instabilities also depends on the angular distributions and favorable lepton numbers. In the regions of neutrino decoupling, conditions are the most favorable for flavor instability due to neutrino self-interactions that can lead to collective effects. These interactions render the flavor evolution problem inherently non-linear, giving rise to a variety of instabilities and oscillation modes. The behavior of neutrinos under these conditions is governed by the quantum kinetic equations (QKEs), which encapsulate both coherent flavor oscillations and inelastic scattering processes \cite{Sigl:1993ctk}.   

A defining feature of flavor evolution in dense media is the emergence of collective effects \cite{Mirizzi2016}. Depending on the dominant physical scale, different regimes can arise—slow collective oscillations that depend on the vacuum mixing parameters \cite{Duan:2010bg, Mirizzi2016}, fast flavor conversions (FFC) driven by electron lepton number (ELN) angular crossings \cite{Sawyer:2005jk, Sawyer2016, Izaguirre:2016gsx, Abbar2019, Capozzi:2020syn, Morinaga2022}, and collisional flavor instabilities (CFI) \cite{Johns2023, Xiong2023, Kato2023} arising from interactions with the dense medium. These effects are deeply sensitive to local conditions, such as the angular structure of neutrino emission and the degree of anisotropy.

While the impact of neutrino heating has been incorporated in state-of-the-art CCSN simulations \cite{Janka:2006fh, Burrows:2020qrp}, flavor-changing processes are often ignored or simplified by assuming flavor conservation. Such simplifications could miss important feedback, since changes in the flavor composition alter neutrino energy and angular spectra, affecting both shock revival and the neutron-to-proton ratio relevant for nucleosynthesis \cite{Tamborra:2017ubu, Abbar2019}. 

Accurately incorporating neutrino flavor evolution in multi-dimensional CCSN simulations presents a major computational challenge. Typically, the numerical solutions of QKEs require orders of magnitude more computational resources compared to the simplified classical transport schemes, such as M1 or the neutrino leakage scheme, or even the full Boltzmann solution. It should be noted that the neutrino transport is one of the most expensive parts of the hydrodynamical simulations of CCSN. Incorporating solutions of QKEs in multi-dimensional simulations is not yet feasible.  

Nevertheless, progress can be made by solving the QKEs in simplified frameworks or localized regions within simulations \cite{Shalgar:2019kzy, Padilla-Gay:2020uxa, Shalgar:2020wcx,  Padilla-Gay:2021haz, Shalgar:2022lvv, Richers:2022dqa, Shalgar:2023aca}. Such approaches allow us to probe the qualitative behavior, investigate the conditions under which various flavor instabilities arise, and assess their effects on neutrino spectra. More importantly, these studies can reveal whether flavor equilibration, where different neutrino species approach similar angular or spectral distributions, is a generic outcome of flavor evolution. However, such simulations do not take into account the feedback of neutrino flavor evolution on the hydrodynamic evolution. Including this feedback is a very challenging task and one of the major goals for the future. 

In this work, we focus on neutrino flavor evolution in spherical symmetry and investigate the robustness of quantum kinetic equation solutions to matter fluctuations during the accretion phase of a core-collapse supernova. Using post-bounce profiles from one-dimensional hydrodynamic simulations of an $18.6 M_\odot$ progenitor \cite{Garching_CCSN_archive}, we introduce both time-independent and time-dependent matter fluctuations to assess their impact on flavor evolution. This result is in agreement with the conclusions of a previous study that looked at the effect of perturbations by adding fluctuations to the neutrino gas itself in an ad hoc manner~\cite{Cornelius:2023eop}. Our results show that the solutions to the QKEs remain remarkably stable and insensitive to such fluctuations. Furthermore, we performed a resolution study using radial grids with 150, 1500, and 15000 bins, and found excellent agreement across all three cases, demonstrating that even relatively coarse radial resolution captures the essential flavor dynamics.

The paper is structured as follows: in Section~\ref{SphericalModel}, we discuss the spherically symmetric physical model. Then in Section~\ref{MatterFluctuations} we study matter fluctuations and the sensitivity of the QKEs to these fluctuations. In particular, we first discuss time-independent fluctuations in Section~\ref{TimeIndependent}. Then, time-dependent fluctuations in Section~\ref{TimeDependent}. We show the results and robustness of the solutions of the QKEs in Section~\ref{Results}. In Section~\ref{SimulationSetup} we describe the simulation setup. In Section~\ref{Resolution}, we present the numerical convergence of the solutions of QKEs for three different radial resolutions. Finally, in Section~\ref{Discussion} we discuss and summarize the main results and future prospects.

\section{Physical Model}\label{SphericalModel}
\subsection{Spherically Symmetric Approximation}
To model neutrino flavor evolution in core-collapse supernovae, we follow \cite{Shalgar:2024gjt} and solve QKEs under the assumption of spherical symmetry. We consider four neutrino flavors: $\nu_e$, $\bar{\nu}_e$, $\nu_x$, $\bar{\nu}_x$, where $\nu_x$ and $\bar{\nu}_x$ represent characteristic heavy neutrino and antineutrino respectively. The neutrino ensemble is described by $2\times 2$ Hermitian density matrices, $\rho$ and $\bar{\rho}$, which are the function of radius $r$, polar angle $\theta$, neutrino energy $E$, and time $t$. The diagonal elements of the density matrix $\rho_{ii}$ represent the flavor-resolved phase-space occupation numbers, and integrating over angle and energy yields the local number density for each flavor.

The QKEs govern the time evolution of neutrino and antineutrino density matrices: 
\begin{eqnarray}
    \bigg( \frac{\partial}{\partial t} + \boldsymbol{v}\cdot \boldsymbol{\nabla}\bigg)\rho = [H, \rho] +i\mathcal{C}[\rho] \, , \label{eq:QKE_rho}\\ 
     \bigg( \frac{\partial}{\partial t} + \boldsymbol{v}\cdot \boldsymbol{\nabla}\bigg)\bar{\rho} = [\bar{H}, \bar{\rho}] +i\bar{\mathcal{C}}[\bar{\rho}] \, , \label{eq:QKE_rhobar}
\end{eqnarray}
where $\boldsymbol{v}$ denotes neutrino propagation direction and has a magnitude equal to the speed of light. Under the assumption of spherical symmetry, the advection term is
\begin{eqnarray}
    \boldsymbol{v} \cdot \boldsymbol{\nabla} = \cos{\theta} \frac{\partial}{\partial r} +\frac{\sin^2\theta}{r} \frac{\partial}{\partial \cos\theta} \, . \label{eq:advection}
\end{eqnarray}

Eqs.~(\ref{eq:QKE_rho}) and (\ref{eq:QKE_rhobar}) capture the advection of neutrinos through the supernova medium as well as their flavor transformation and collisional processes. It should be noted that in our formalism, the advective term does not include the relativistic effects. The flavor conversion is described by the commutator terms, $[H, \rho]$ and $[\bar{H}, \bar{\rho}]$, where the Hamiltonian is $H=H_{\rm vac} + H_{\nu \nu}$ for neutrinos and $\bar{H}=-H_{\rm vac} + H_{\nu\nu}$ for antineutrinos. Rather than explicitly including the matter potential, we approximate its flavor-suppressing effect by using the vacuum mixing angle. The vacuum Hamiltonian is
\begin{equation}
    H_{\rm vac} (E)=\frac{\omega_{\rm vac}}{2} \begin{pmatrix}
    -\cos2\vartheta_V & \sin2\vartheta_V \\
    \sin 2\vartheta_V & \cos2\vartheta_V
    \end{pmatrix} \, , \label{eq:Hamiltonian_vacuum}
\end{equation}
where 
\begin{equation}
    \omega_{\rm vac} = \frac{\Delta m^2}{2E} \, .
\end{equation}
We adopt $\Delta m^2 = 2.5 \times 10^{-3} ~eV^2$ and effective mixing angle $\vartheta_V = 10^{-3} ~\rm rad$. 

Neutrino-neutrino interactions are incorporated via the self-interacting Hamiltonian
\begin{eqnarray}
      H_{\nu \nu}(r, \cos \theta, t) = \sqrt{2} G_{F}\zeta \int_{-1}^1 \int_0^\infty & &[\rho(r, \cos\theta', E, t) - \bar{\rho}(r, \cos\theta', E, t)] \\ \nonumber
    &\times& (1-\cos\theta \cos \theta') dE d\cos\theta' \, ,
\end{eqnarray}
where $G_F$ is the Fermi constant. To accelerate the simulation towards a quasi-steady state while retaining its qualitative features, similar to \cite{Nagakura2022}, we rescale the interaction strength by a factor $\zeta=10^{-3}$. The integral over azimuthal angles results in a factor of $2\pi$, which we absorb in the normalization of the density matrices.

On the right side of Eqs.~(\ref{eq:QKE_rho}) and (\ref{eq:QKE_rhobar}), we have the collision term, which includes neutrino emission, absorption, and elastic scattering processes. Emission and absorption arise from both thermal pair processes and charged-current reactions, with the latter affecting only electron-type flavors. Elastic scattering with nucleons alters neutrino directions but conserves energy. The collision term has the form:
\begin{equation}
    \mathcal{C}[\rho]=\mathcal{C}_{\rm emit} (E) - \mathcal{C}_{\rm abs} (E) ~\odot ~\rho + \int d\cos\theta' \cos{\theta'}[\rho(\cos\theta')-\rho(\cos\theta)] \mathcal{C}_{\rm dir-ch}(E) \, ,
\end{equation}
where $\odot$ denotes element-wise multiplication of matrices. The emission and absorption rates satisfy detailed balance through Kirchhoff’s law:
\begin{equation}
    \frac{\mathcal{C}_{\rm emit}(E)}{\mathcal{C}_{\rm abs}(E)}= f_{\nu_i}^{\rm FD}(E) \, ,
\end{equation}
with $f_{\nu_i}^{\rm FD}(E)$ the Fermi-Dirac distribution for flavor $i$. 
Our treatment follows the multi-energy framework of \cite{Shalgar:2023aca}, with interaction cross sections taken from \cite{OConnor:2014sgn}. The collision terms require thermodynamic variables from a core-collapse supernova simulation; in this work, we use those provided by \cite{Garching_CCSN_archive}. For all the simulations in this paper, we use the profiles for post-bounce time of 1 second, as it exhibits fast flavor instability.

 \begin{figure}
     \centering
     \includegraphics[width=0.8\textwidth]{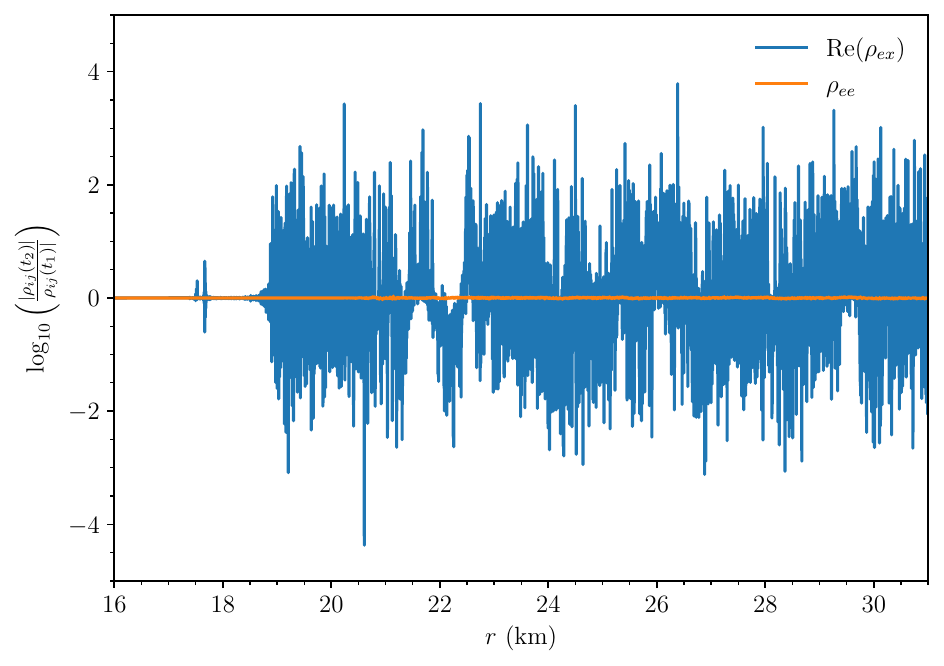}
     \caption{Spatial changes between two short times in the QKE evolution. Plotted is $\log_{10}\!\bigl(|\rho_{ij}(t_2)/\rho_{ij}(t_1)|\bigr)$ versus radius for $t_1=0.9\times10^{-4}\,\mathrm{s}$ and $t_2=1.0\times10^{-4}\,\mathrm{s}$ (over $r\approx16\text{--}31\ \mathrm{km}$). The orange curve shows the diagonal element $\rho_{ee}$, which remains near zero on this log-ratio scale, indicating $\rho_{ee}(t_2)\simeq\rho_{ee}(t_1)$. The blue curve shows $\mathrm{Re}(\rho_{ex})$, which exhibits large, spatially intermittent excursions spanning several orders of magnitude; this demonstrates that off-diagonal coherences continue to evolve on short timescales even after the diagonal occupations have stabilized. Data are from a QKE run initialized with the Boltzmann steady state and evolved until the off-diagonal amplitudes reached a quasi-steady value; the phase and fine structure of those off-diagonals remain time-dependent.}
     \label{fig:quasi_steady_state}
 \end{figure}

\begin{figure}
    \centering
    \includegraphics[width=\linewidth]{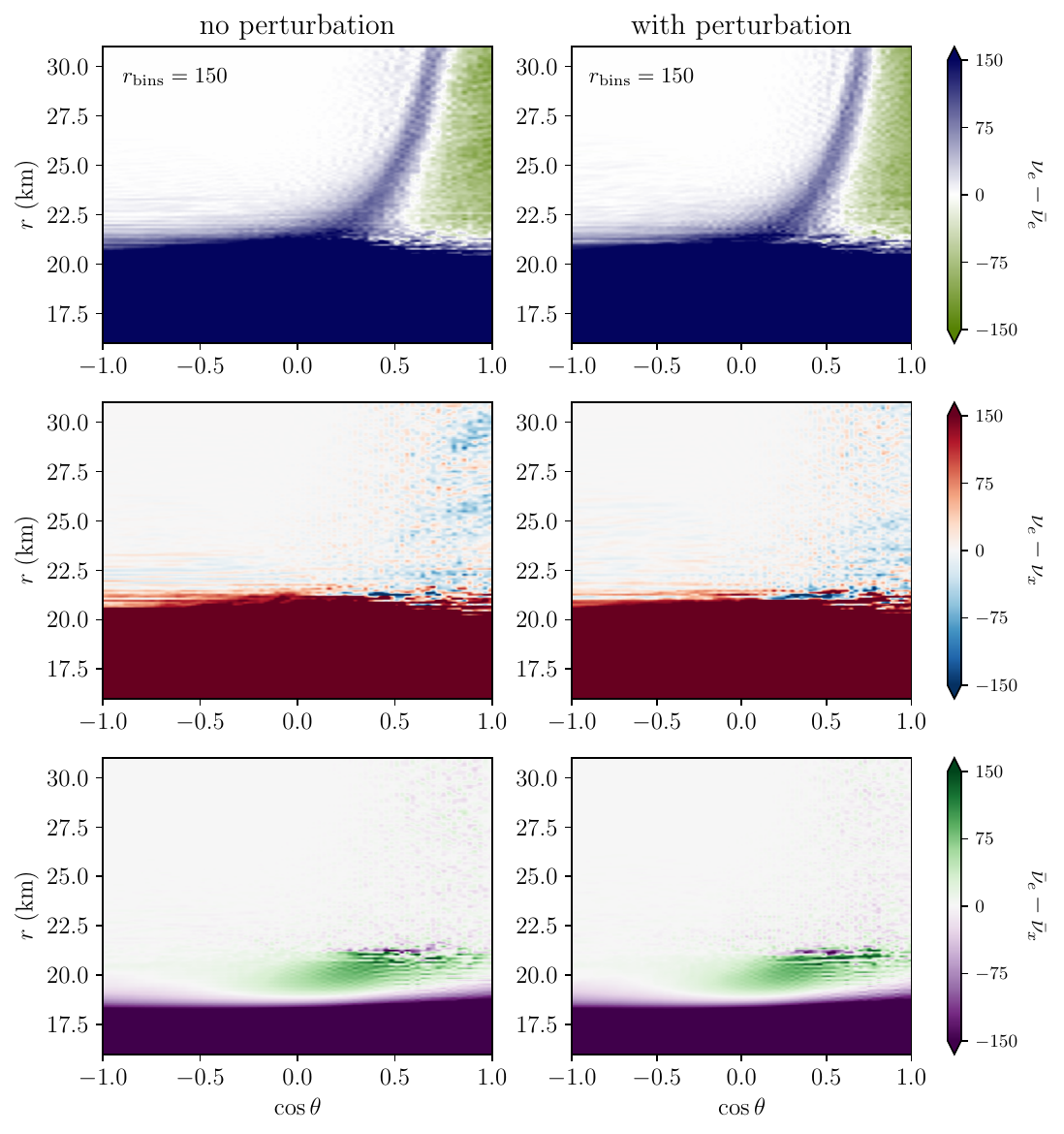}
    \caption{Heatmaps of the energy-integrated angular distributions of neutrino flavor asymmetries for \( r_{\mathrm{bins}} = 150 \). The left column shows results without perturbations, while the right column includes time-independent perturbations to the collision terms. From top to bottom, the rows correspond to \( \nu_e - \bar{\nu}_e \), \( \nu_e - \nu_x \), and \( \bar{\nu}_e - \bar{\nu}_x \), respectively. The perturbations induce no visible modifications to the angular structure.}
    \label{fig:ang_dist150}
\end{figure}

\begin{figure}
    \centering
    \includegraphics[width=\linewidth]{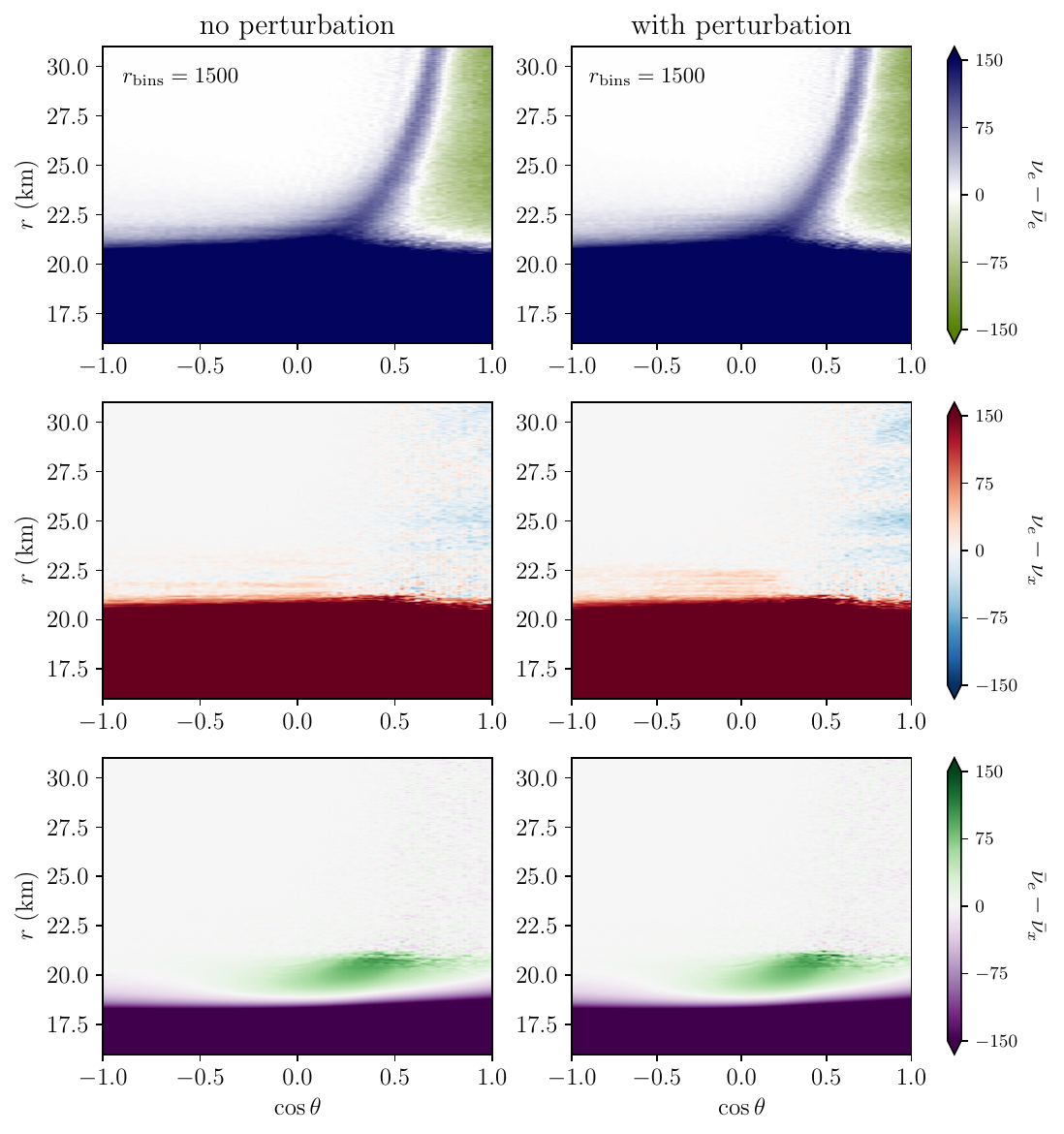}
    \caption{Same as Fig.~\ref{fig:ang_dist150} but for \( r_{\mathrm{bins}} = 1500 \). The angular distributions show negligible differences compared to the \( r_{\mathrm{bins}} = 150 \) case, indicating that the solution is well converged with respect to radial resolution. The impact of the perturbations remains consistent across resolutions.}
    \label{fig:ang_dist1500}
\end{figure}

\begin{figure}
    \centering
    \includegraphics[width=\linewidth]{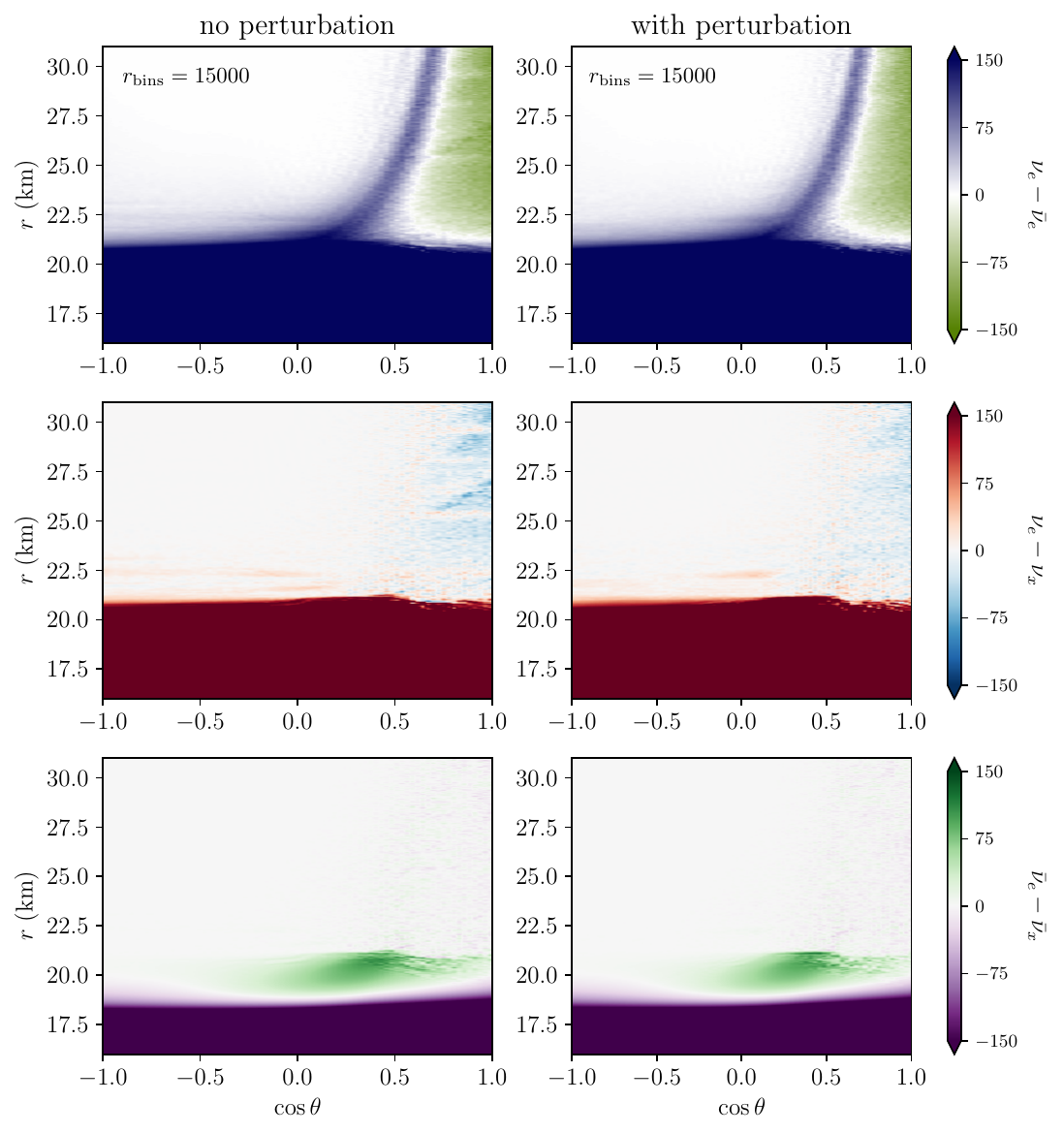}
    \caption{Same as Fig.~\ref{fig:ang_dist150} but for \( r_{\mathrm{bins}} = 15000 \). Like in Fig.~\ref{fig:ang_dist1500}, the angular distributions show negligible differences compared to the \( r_{\mathrm{bins}} = 150 \) case.}
    \label{fig:ang_dist15000}
\end{figure}

\begin{figure}
    \centering
    \includegraphics[width=\linewidth]{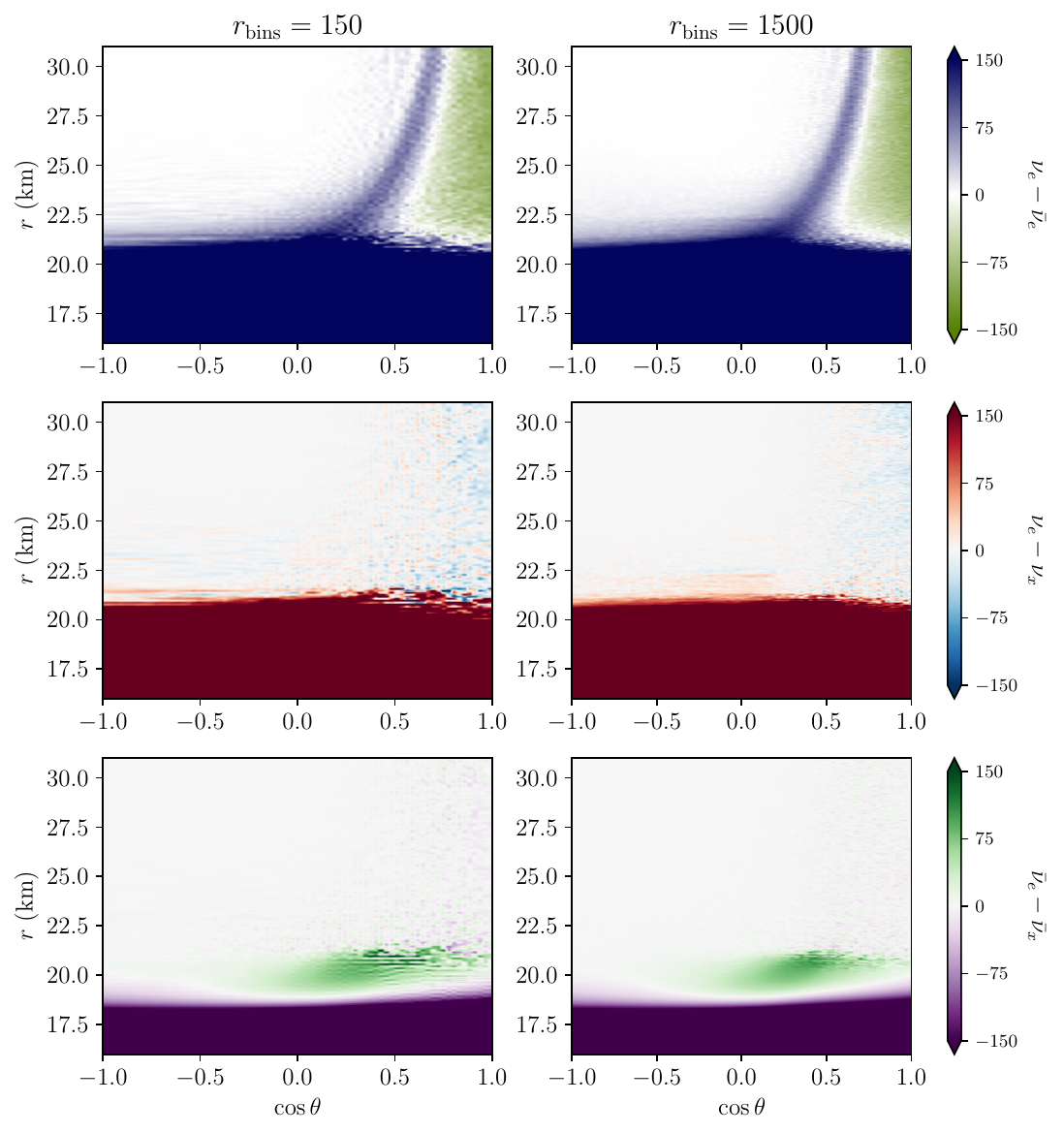}
    \caption{Angular distributions of neutrino flavor differences with time-dependent perturbations applied to the collision terms. The left column shows results for \( r_{\mathrm{bins}} = 150 \), and the right column for \( r_{\mathrm{bins}} = 1500 \). Each row corresponds to a different flavor combination: \( \nu_e - \bar{\nu}_e \), \( \nu_e - \nu_x \), and \( \bar{\nu}_e - \bar{\nu}_x \) (top to bottom). The overall structure remains consistent across resolutions, confirming that the results are converged even with rapidly varying matter fluctuations.}
    \label{fig:enter-label}
\end{figure}

\begin{figure}
    \centering
    \includegraphics[width=\linewidth]{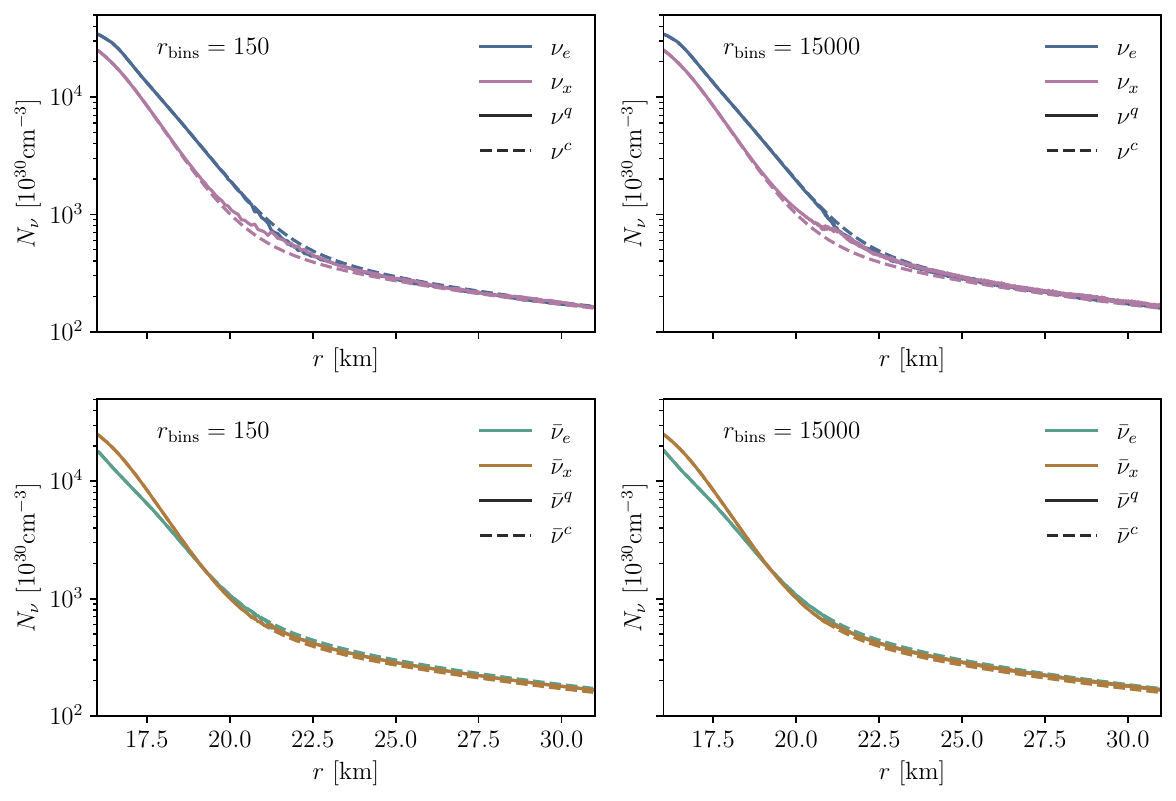}
    \caption{Radial profiles of angle-integrated neutrino number densities for different flavors, comparing classical (\(N_\nu^{\mathrm{c}}\), dashed) and quantum (\(N_\nu^{\mathrm{q}}\), solid) solutions in the case of no perturbation. The top row shows results for \(\nu_e\) and \(\nu_x\), while the bottom row shows \(\bar{\nu}_e\) and \(\bar{\nu}_x\). Left and right columns correspond to radial resolutions \(r_{\mathrm{bins}} = 150\) and \(r_{\mathrm{bins}} = 15000\), respectively. The close agreement between the two resolutions confirms numerical convergence, with quantum corrections introducing small but consistent deviations from classical results.}
    \label{fig:numb_dens_np}
\end{figure}

\begin{figure}
    \centering
    \includegraphics[width=\linewidth]{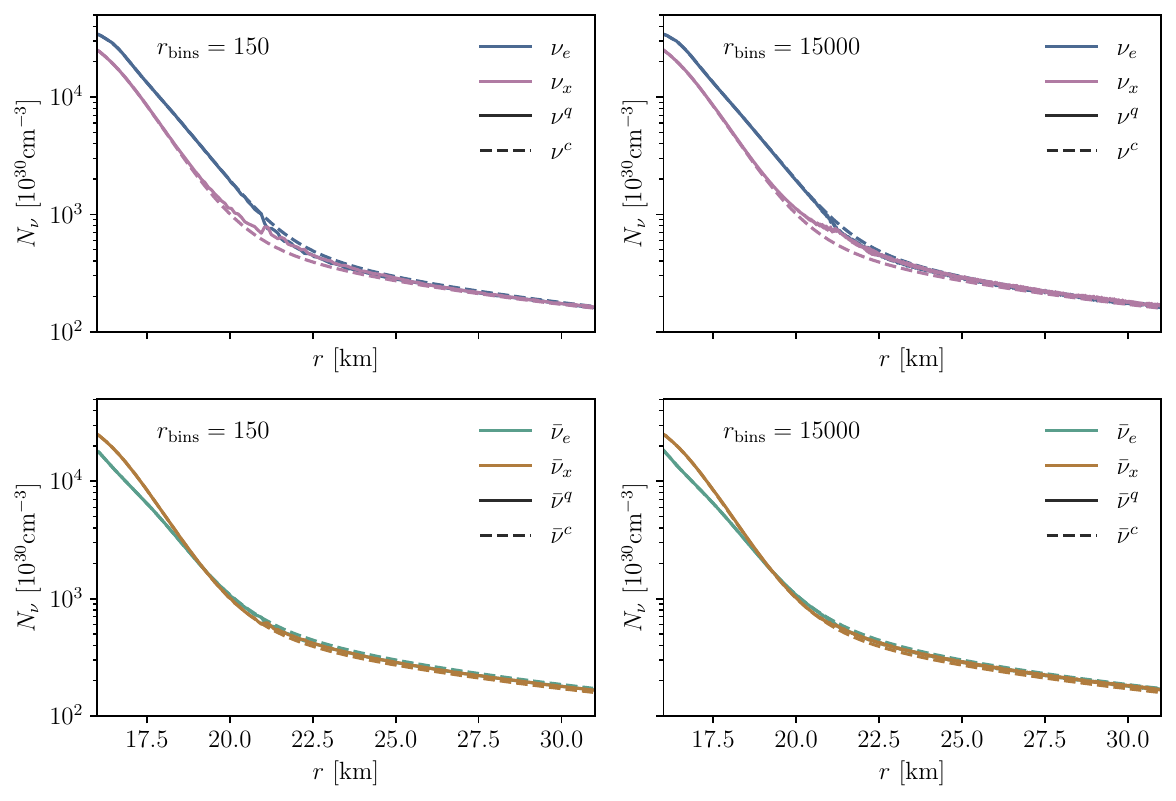}
    \caption{Same as Fig.~\ref{fig:numb_dens_np} now in the case of time-independent matter fluctuations.}
    \label{fig:numb_dens_cp}
\end{figure}

\begin{figure}
    \centering
    \includegraphics[width=\linewidth]{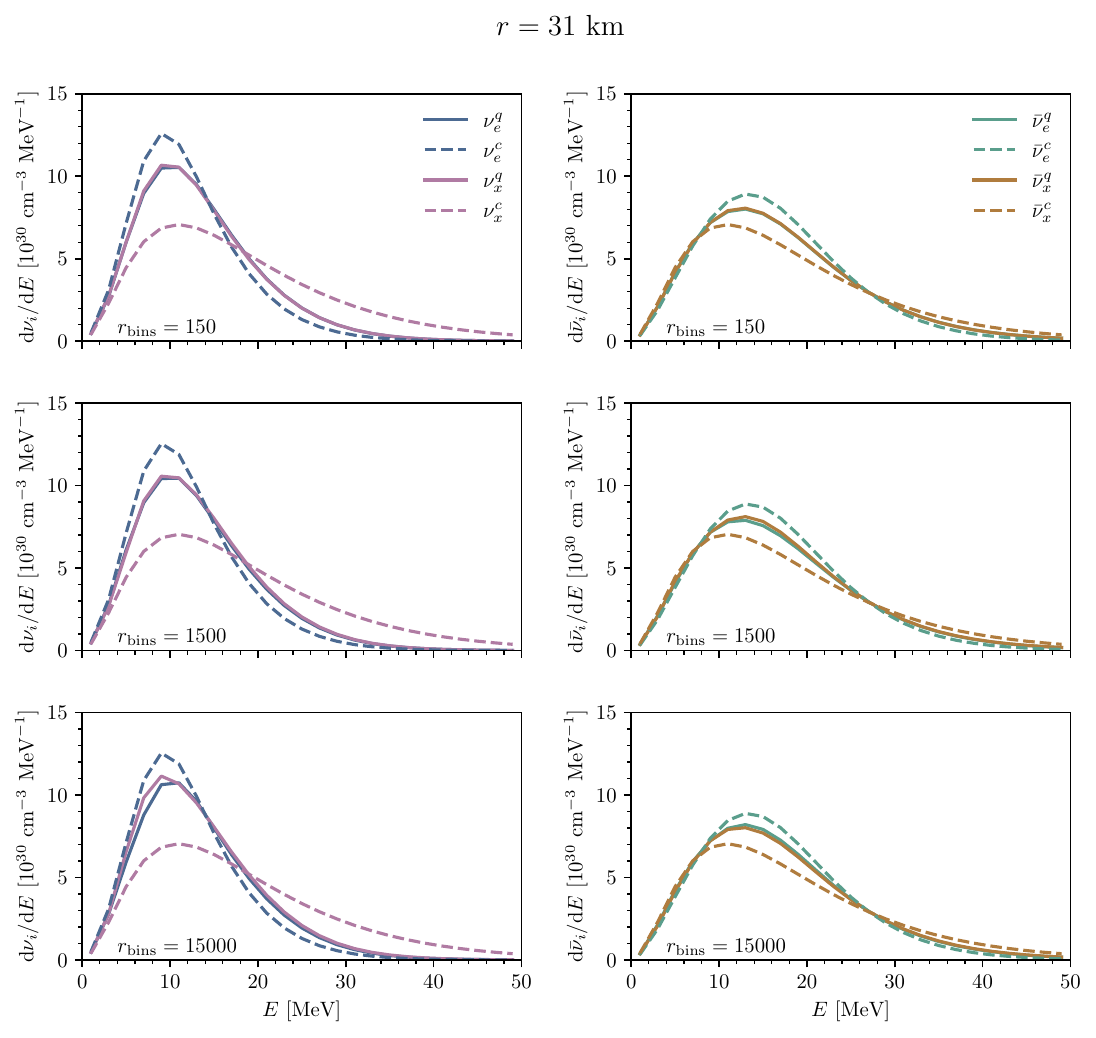}
    \caption{Unperturbed energy spectra \(d\nu/dE\) for \(\nu_e\), \(\nu_x\), \(\bar{\nu}_e\), and \(\bar{\nu}_x\) at a fixed radius of \(r = 31\) km. Solid and dashed lines show the quantum (\(q\)) and classical (\(c\)) results, respectively. The top, middle, and bottom rows correspond to radial resolutions \(r_{\mathrm{bins}} = 150\), 1500, and 15000. The left and right columns show neutrino and antineutrino distributions, respectively. Quantum effects alter the spectral shape slightly, but the results are numerically converged across different resolutions.
}
    \label{fig:en_spectrum_np}
\end{figure}

\begin{figure}
    \centering
    \includegraphics[width=\linewidth]{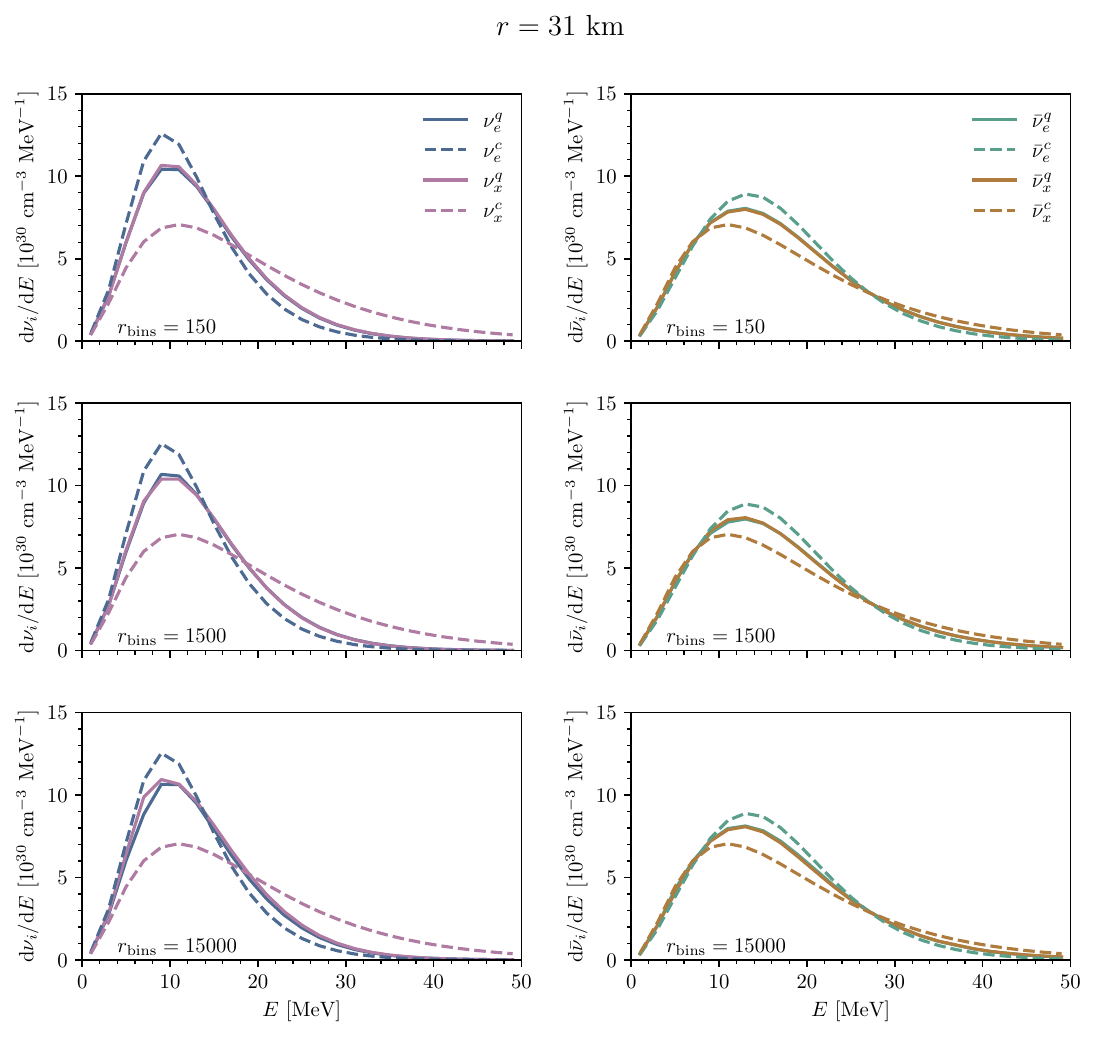}
    \caption{Same as Fig.~\ref{fig:en_spectrum_np} but in the case of constant matter fluctuations.}
    \label{fig:en_spectrum_cp}
\end{figure}

\begin{figure}
    \centering
    \includegraphics[width=\linewidth]{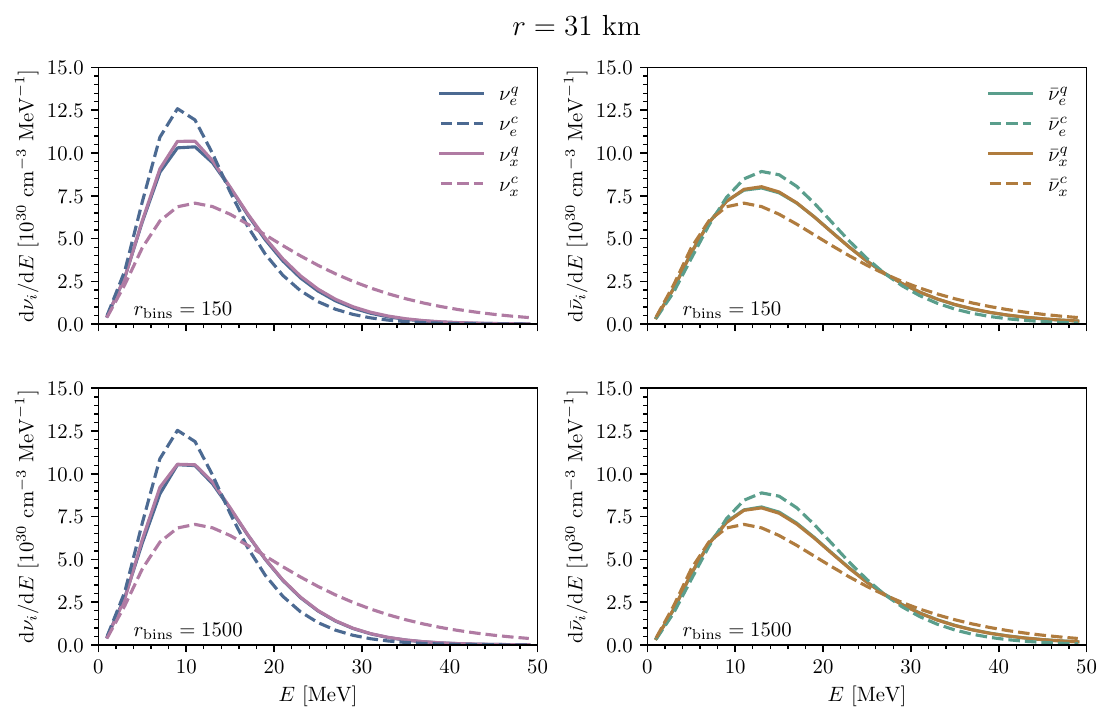}
    \caption{Same as Fig.~\ref{fig:en_spectrum_np} in the case of time-dependent fluctuations.}
    \label{fig:enter-label}
\end{figure}
\subsection{Matter Fluctuations}\label{MatterFluctuations}
In realistic astrophysical environments, the matter background is subject to fluctuations arising from turbulence, convection, and shock propagation. These variations in density and temperature lead to corresponding changes in neutrino interaction rates, since the collision term depends directly on local fluid properties. To explore how such small-scale inhomogeneities affect neutrino flavor evolution, we introduce stochastic perturbations to the collision term in the quantum kinetic framework.

We model these matter fluctuations by rescaling each component of the collision term \( \mathcal{C}[\rho] \) with a random factor. For each collision process, the perturbed term takes the form
\begin{equation}
    \mathcal{C}'[\rho] = (1 + n) \, \mathcal{C}[\rho],
\end{equation}
where \( n \) is a random number drawn from a distribution with zero mean and standard deviation \( \sigma \) for each radial bin and energy bin. In this work, we typically choose \( \sigma = 0.01 \), corresponding to 1\% fluctuations in the interaction rates. This rescaling can be viewed as a proxy for unresolved matter inhomogeneities that modulate the effective opacity and emissivity. Since the origin of the fluctuations is assumed to be turbulent matter motion, we apply the same perturbation factor across all angular bins, treating them as isotropic in angle space.

Two limiting cases are considered: time-independent fluctuations that are fixed throughout the evolution, and time-dependent fluctuations that vary on short timescales. These are described below.

\subsubsection{Time-independent Fluctuations}\label{TimeIndependent}

Time-independent fluctuations correspond to the static perturbation field introduced at the beginning of the simulation and held constant thereafter. 
For each spatial and energy grid points, we generate a random number $n_i$ and rescale the corresponding collision term as 
\begin{equation}
    \mathcal{C}'_i[\rho] = (1 + n_i) \, \mathcal{C}_i[\rho] \, .
\end{equation}

This models a "frozen" background of matter density inhomogeneities, such as may arise from a snapshot of turbulent flow. While the fluctuations themselves don't vary in time, they break spatial symmetry 
and can favor flavor instabilities that develop dynamically.

\subsubsection{Time-dependent Fluctuations}\label{TimeDependent}

In the time-dependent case, the fluctuation field varies stochastically in time. At regular intervals $\Delta t$, each component of the collision term is rescaled using a freshly drawn perturbation:
\begin{equation}
       \mathcal{C}'_i[\rho(t)] = (1 + n_i(t)) \, \mathcal{C}_i[\rho(t)] \, ,
\end{equation}
where $n_i(t)$ is a new random number drawn independently at a regular interval. In our simulation, we update perturbations every $\Delta t = 10^{-8} ~\mathrm{s}$, a timescale chosen to be comparable to the characteristic timescale of flavor evolution. While this value is somewhat extreme, it allows us to probe the response of the system to rapid fluctuations in the matter background and to test the sensitivity of the flavor dynamics to temporal variability in the collision processes.

\section{Simulation Setup}\label{SimulationSetup}
In this section, we describe the numerical setup used for the simulation results presented in this paper. Eqs.~(\ref{eq:QKE_rho}) and (\ref{eq:QKE_rhobar}) are partial differential equations that need to be discretized to be solved numerically. We discretize the QKEs using 75 angle bins in the polar angle that are equally spaced along $\cos\theta$. We use 25 equally spaced energy bins with an energy range [0,50] MeV, and the number of radial bins varies depending on the configuration. The number of radial bins, $r_{\textrm{bins}}$ used varies from 150 to 15000. 

The advective term on the left-hand-side of Eqs.~(\ref{eq:QKE_rho}) and (\ref{eq:QKE_rhobar}) contains derivatives with respect to $r$ and $\cos\theta$, which we calculate using the central difference method. Since the $r$ and $\cos\theta$ bins are equally spaced, the central difference method results in errors that are second order in bin size. 

The thermodynamic background for our quantum kinetic equation calculations is provided by a one-dimensional core-collapse supernova simulation of an $18.6 M_\odot$ progenitor with an SFHo EOS and a neutron star that has a gravitational mass of 1.4 $M_{\odot}$ \cite{Garching_CCSN_archive}. From this simulation, we extract radial profiles of density, temperature, and electron fraction, which are required to evaluate the collision terms and interaction rates. We focus on conditions at $1~\rm s$ after core bounce. This time is of particular interest because fast flavor instabilities are expected to be especially prominent: by this time, the proto-neutron star has cooled and contracted significantly, the neutrino decoupling region has settled into a quasi-steady configuration, and large ELN angular crossings can develop in the semi-transparent layers. These features make the post-bounce 1-second snapshot a favorable setting to study the sensitivity of QKE evolution to small-scale perturbations in the collision terms.

In order to study the flavor evolution, we follow the procedure proposed in ~\cite{Shalgar:2023aca}, and first solve the equations of motion without the flavor evolution terms. Setting the flavor evolution terms to zero converts Eqs.~(\ref{eq:QKE_rho}) and (\ref{eq:QKE_rhobar}) to Boltzmann equations, which we evolve until a steady state is reached. This solution is independent of the configuration used at the start of the simulation, but we find that it is efficient to start the simulation with the Fermi-Dirac distribution. The temporal evolution is carried out using the multistep VCABM3 algorithm with adaptive step size from \texttt{DifferentialEquations.jl} package in Julia~\cite{rackauckas2017differentialequations}. 

After the steady state solution of the Boltzmann equation is found, it is used as the initial condition for the solution of the QKEs, which is obtained by switching on the flavor evolution term in 
 Eqs.~(\ref{eq:QKE_rho}) and (\ref{eq:QKE_rhobar}). It should be noted that this two-step procedure is not necessary but convenient from a numerical point of view for two reasons. Firstly, the solutions of the QKEs are not very far from the solution of the Boltzmann equation, especially in the very high-density region, which prevents large time derivatives. Secondly, it is a convenient way of solving the QKEs as the effect of flavor evolution can be easily disentangled from the Boltzmann solution by looking at the difference between the two solutions. The QKEs are numerically evolved until a quasi-steady state is reached. The quasi-steady state implies that the diagonal components of the density matrices and the absolute value of the off-diagonal components reach a steady state, but the phase of the off-diagonal components never reaches a steady state. Due to this peculiarity, the system of QKEs discussed in this paper cannot be thought of as a boundary value problem which can be solved by requiring that $[H^{\textrm{eq}},\rho^{\textrm{eq}}] = 0$. 

Fig.~\ref{fig:quasi_steady_state} displays the radial pattern of short-time changes in the
density-matrix components, quantified by
\(\log_{10}\!\bigl(|\rho_{ij}(t_2)/\rho_{ij}(t_1)|\bigr)\) for \(t_1=0.9\times10^{-4}\,\mathrm{s}\)
and \(t_2=1.0\times10^{-4}\,\mathrm{s}\) (covering \(r\approx16\text{--}31\ \mathrm{km}\)).
On this scale, the diagonal electron occupation, \(\rho_{ee}\), shows negligible variation between
the two times, indicating that the occupations have effectively settled by \(t_1\). In contrast,
the real part of the off-diagonal element, \(\mathrm{Re}\,\rho_{ex}\), exhibits large, localized deviations across the radius: coherences change by many orders of magnitude in a
spatially intermittent fashion. Taken together, these results demonstrate a separation of
behaviour—populations reach a quasi-steady state on the short timescale considered, while
coherences retain substantial dynamical activity (in amplitude fine structure and phase). This
separation makes clear that capturing the true QKE steady behavior requires explicit time
integration of the evolution equations rather than solving a time-independent commutator
constraint. While \(\rho_{ee}\) has reached the steady state by \(t_1=0.9\times10^{-4}\,\mathrm{s}\), \(\mathrm{Re}\,\rho_{ex}\)
continues to evolve up to \(t_2=1.0\times10^{-4}\,\mathrm{s}\); therefore, the QKE quasi-steady state is intrinsically time-dependent and must be obtained through dynamical evolution.

 In numerical simulations that do not have any perturbations, the collision terms are the same irrespective of whether we solve the Boltzmann equations or the QKEs. However, in the case of QKEs with perturbation, we perturb the collision terms while solving the QKEs. 

 We implement perturbations to the collision terms by rescaling all the collision terms by $\sim 1$\%. This is done by multiplying each collision term by $(1+n)$, where $n$ is a random number chosen from a random distribution with a standard deviation of 0.01. We assume that the perturbations are independent of the angle. In the case of perturbations that we call `time independent', the perturbations are applied at the beginning of the numerical simulation that includes the flavor conversions, and the perturbations are not changed with time until the quasi-steady state is reached. For simulations labeled as having time-dependent perturbations, the procedure used is similar to the one described above, but the random perturbations are updated every $10^{-8}$~s.

\subsection{Resolution Study and Numerical Convergence}\label{Resolution}

In addition to the baseline simulation setup, we perform a resolution study to assess the numerical convergence of our results with respect to radial discretization. Specifically, we vary the number of radial grid points, $r_{\mathrm{bins}}$, using values of $150$, $1500$, and $15000$ while keeping all the other numerical and physical parameters fixed. This allows us to test the sensitivity of flavor evolution to the radial resolution used in evaluating collision terms and integrating the QKEs. 
We find that the results obtained with $r_{\mathrm{ bins}}=1500$ and $r_{\mathrm{bins}}=15000$ are nearly identical to the results obtained with the coarser grid $r_{\mathrm{bins}}=150$, indicating that the solution converged at this level of resolution. This demonstrates that even the coarser grid reproduces the key features of flavor evolution with high accuracy.

This resolution study shows that our simulations are numerically robust and that $r_{\mathrm{bins}}=150$ provides a sufficient level of accuracy for the purposes of this study.

\section{Results}\label{Results}

We now present the results of our simulations, comparing classical and quantum transport solutions across a range of conditions. We focus on how angular distributions, number densities, and energy spectra of neutrinos vary under different perturbation schemes and radial resolutions.

\subsection{Angular Distributions and Convergence}

Figs.~\ref{fig:ang_dist150}, \ref{fig:ang_dist1500}, and \ref{fig:ang_dist15000} show the energy-integrated angular distributions of neutrino flavor asymmetries for \( r_{\mathrm{bins}} = 150 \), 1500, and 15000, respectively. The left columns show the baseline case with no perturbations, while the right columns include time-independent perturbations to the collision terms. The differences between perturbed and unperturbed runs are negligible, indicating that static matter fluctuations do not strongly modify the angular structure.

In all three resolutions, the flavor asymmetry patterns remain qualitatively and quantitatively consistent, suggesting excellent convergence of the angular distributions. Increasing radial resolution does not introduce significant changes, confirming the numerical stability of the solution.

Note that we also performed the same analysis with a time-independent case, but with a larger perturbation of $\sim 10\%$ to the collision terms. However, the results and main conclusions are the same as for the case of $\sim 1\%$ case. Therefore, we do not show those results in this paper.

Fig.~\ref{fig:enter-label} extends this analysis to include time-dependent fluctuations in the matter profile. Even in the presence of rapidly varying perturbations, the angular distributions remain stable, and the structure is preserved across different resolutions. This confirms that the system is robust to time-dependent fluctuations in the matter background.

\subsection{Angle-Integrated Number Densities}

The angle-integrated neutrino number densities are shown in Figs.~\ref{fig:numb_dens_np} and \ref{fig:numb_dens_cp}, corresponding to the unperturbed case and the case with constant fluctuations, respectively. In each plot, classical and quantum results are compared for multiple neutrino species, and resolutions \( r_{\mathrm{bins}} = 150 \) and 15000 are shown.

The results demonstrate that quantum corrections induce small but consistent deviations in the number densities compared to the classical case, particularly in the outer regions. However, the agreement between the two resolutions is excellent, once again confirming numerical convergence. The introduction of constant fluctuations does not significantly alter this behavior, showing that the global impact of static perturbations is limited.

\subsection{Energy Spectra}

The differential energy spectra \( d\nu/dE \) at a fixed radius of 31 km are displayed in Figs.~\ref{fig:en_spectrum_np}, \ref{fig:en_spectrum_cp}, and \ref{fig:enter-label}, corresponding to the unperturbed, constant perturbation, and time-dependent perturbation cases, respectively. Each figure shows results for all neutrino flavors, comparing quantum and classical solutions across resolutions \( r_{\mathrm{bins}} = 150 \), 1500, and 15000.

The spectral shapes are nearly identical between the classical and quantum cases, with quantum effects inducing subtle but systematic shifts in the spectral tails. These differences are more prominent at higher energies and are well resolved even at moderate resolution, as evidenced by the excellent agreement between the different \( r_{\mathrm{bins}} \) values.

Overall, the spectra are robust against both static and dynamic perturbations, and all results show strong numerical convergence, providing confidence in the stability and physical reliability of the simulations.

\section{Discussion and Conclusions}\label{Discussion}

We have studied the robustness of the numerical solutions of QKEs in a spherical geometry by introducing perturbations to the collision terms, which are the only physics inputs in the problem other than fundamental constants. We find that the numerical solutions of the QKEs are virtually unaffected by the perturbations. In our analysis, we use two different approaches to introduce perturbations. In the first, the perturbations are time-dependent, evolving dynamically as the system progresses. In the second, they are time-independent, remaining fixed throughout the evolution and thereby influencing the system in a steady, unchanging manner. In both cases, the perturbations lead to no visible impact.

All simulations point to the same conclusion: the numerical results of QKE evolution are highly robust. They show little sensitivity to small-scale perturbations in the collision terms.
It is possible to understand this robustness by considering the mean-free-path of the neutrinos in relation to the length scale of the matter perturbation. If the scale of the matter perturbation is smaller than the mean-free-path of the neutrinos, then the neutrinos cannot see the perturbations and hence the solution of the QKEs is not dependent on these small-scale perturbations.

The motivation for this study is rooted in the nonlinear and potentially chaotic nature of the QKEs~\cite{Hansen:2014paa, Urquilla:2024bvf}. It may be argued that due to the nonlinear nature of the QKEs, the solutions of the QKEs are very sensitive to the input parameters. The collision terms are the only input parameters in the problem and are dependent on the matter density, which is usually assumed to be smooth. Although a direct coupling between perturbations in the matter density and the flavor evolution has been speculated in the past~\cite{Fiorillo:2024bzm}, ours is the first paper to investigate this in detail.

We have scaled the collision terms of random numbers that are normally distributed with a standard deviation of 1\%. The collision terms are modified for each radial bin, thus making the resolution of the simulation a proxy for the length scale of the perturbation. In simulations that we call time-independent, we use the same perturbation throughout the simulation, whereas in the case of time-dependent perturbation, we generate new random numbers every $10^{-8}$ seconds. We have performed the simulations with time-independent perturbations with resolutions of 150, 1500, and 15000 radial bins, which correspond to length scales of perturbation that are 100, 10, and 1 meters, respectively. We have performed the same analysis with time-dependent perturbations with 150 and 1500 radial bins. 

In this paper, we have adopted a simplified setup in which the resolution of the simulation sets the minimal length scale of the imposed matter density fluctuations. In realistic supernova environments, however, turbulence can generate fluctuations across a broad range of length scales and amplitudes. Our simplified approach, therefore, serves as a first step, demonstrating the robustness of QKE solutions against small-scale perturbations. A more comprehensive investigation of turbulence-induced fluctuations spanning multiple scales will be an important direction for future work.

 Despite the nonlinear and potentially chaotic nature of the flavor evolution, the robustness of the solution of the QKEs is not surprising. However, we have verified it with a thorough investigation. We thus conclude that the kind of matter density perturbation does not affect neutrino flavor evolution in dense astrophysical environments such as core-collapse supernovae.


\acknowledgments

The authors thank Thomas Janka for sharing the outputs of the hydrodynamical simulations of his group and for reading the draft of the paper and providing insightful comments and feedback. The Tycho supercomputer hosted at the SCIENCE HPC Center at the University of Copenhagen was used to support the numerical simulations presented in this work. This project has received support from the Villum Foundation (Project No. 13164, PI: I. Tamborra), the Danmarks Frie Forskningsfond (Project No. 8049-00038B, PI: I. Tamborra), and the European Union (ERC, ANET, Project No. 101087058). Views and opinions expressed are those of the authors
only and do not necessarily reflect those of the European Union or the European Research Council. Neither the European Union nor the granting authority can be held responsible for them.


\bibliographystyle{JHEP}

\bibliography{references.bib}

\end{document}